\documentclass[iop]{emulateapj}
\shorttitle{TR Penumbral Bright Dots}
\shortauthors{DENG ET AL.}
\submitted{Received 2016 --; accepted 2016 --; published 2016 --}
\journalinfo{Draft version \today}
\usepackage[breaklinks=true,setpagesize=false,bookmarks=false,colorlinks=true,linkcolor=blue,citecolor=blue,urlcolor=blue]{hyperref}

\newcommand{\CII}{\ion{C}{2}}
\newcommand{\CaII}{\ion{Ca}{2}}
\newcommand{\CaIIH}{\ion{Ca}{2}~H}

\newcommand{\SiIV}{\ion{Si}{4}}

\newcommand{\MgII}{\ion{Mg}{2}}
\newcommand{\kms}{km$\,$s$^{-1}$}

\newcommand{\ha}{H$\alpha$}

\begin{document}

\title{Multi-wavelength Study of Transition Region Penumbral Subarcsecond Bright Dots Using IRIS and NST}

\author{NA DENG\altaffilmark{1,2}, VASYL YURCHYSHYN\altaffilmark{2,3}, HUI TIAN\altaffilmark{4}, LUCIA KLEINT\altaffilmark{5}, CHANG LIU\altaffilmark{1,2}, YAN XU\altaffilmark{1,2}, AND HAIMIN WANG\altaffilmark{1,2}}
\affil{$^1$~Space Weather Research Laboratory, New Jersey Institute of Technology, University Heights, Newark, NJ 07102-1982, USA; \href{mailto:na.deng@njit.edu}{na.deng@njit.edu}}
\affil{$^2$~Big Bear Solar Observatory, New Jersey Institute of Technology, 40386 North Shore Lane, Big Bear City, CA 92314-9672, USA;}
\affil{$^3$~Korea Astronomy and Space Science Institute, Daejeon, 305-348, South Korea;}
\affil{$^4$~School of Earth and Space Sciences, Peking University, Beijing, 100871, China;}
\affil{$^5$~University of Applied Sciences and Arts Northwestern Switzerland, Bahnhofstr. 6, 5210 Windisch, Switzerland}

\begin{abstract}
Using high-resolution transition region (TR) observations taken by the Interface Region Imaging Spectrograph (IRIS) mission, \citet{Tian+etal.BD.2014ApJ...790L..29T} revealed numerous short-lived subarcsecond bright dots (BDs) above sunspots (mostly located in the penumbrae), which indicate yet unexplained small-scale energy releases. Moreover, whether these subarcsecond TR brightenings have any signature in the lower atmosphere and how they are formed are still not fully resolved. This paper presents a multi-wavelength study of the TR penumbral BDs using a coordinated observation of a near disk-center sunspot with IRIS and the 1.6~m New Solar Telescope (NST) at the Big Bear Solar Observatory. NST provides high-resolution chromospheric and photospheric observations with narrow-band \ha\ imaging spectroscopy and broad-band TiO images, respectively, complementary to IRIS TR observations. A total of 2692 TR penumbral BDs are identified from a 37-minute time series of IRIS 1400~\AA\ slitjaw images. Their locations tend to be associated more with downflowing and darker fibrils in the chromosphere, and weakly associated with bright penumbral features in the photosphere. However, temporal evolution analyses of the BDs show that there is no consistent and convincing brightening response in the chromosphere. These results are compatible with a formation mechanism of the TR penumbral BDs by falling plasma from coronal heights along more vertical and dense magnetic loops. The BDs may also be produced by small-scale impulsive magnetic reconnection taking place sufficiently high in the atmosphere that has no energy release in the chromosphere.

\end{abstract}


\section{INTRODUCTION}

While sunspots have been observed for a long time, their fine structure and vertical extension are only revealed in recent decades by high-resolution and multi-wavelength (beyond visible) observations. Sunspots are concentrations of strong magnetic fields emerged from the convection zone. The magnetic fields are more vertical in the umbra and more inclined in the penumbra. In the photosphere, penumbral magnetic fields are observed to consist of two components, i.e., a more horizontal component carrying most of the outward Evershed flows is embedded in a more vertical background component \citep[the so-called ``uncombed'' or ``interlocking-comb'' structure, e.g., ][]{Solanki+Montavon1993A&A...275..283S, Lites+etal1993ApJ...418..928L, BellotRubio+etal2004A&A...427..319B, Ichimoto+etal2007PASJ...59S.593I, Deng+etal2010ApJ...719..385D, Borrero+Ichimoto2011LRSP....8....4B}. Recent high-resolution observations \citep[e.g.,][]{Ichimoto2010mcia.conf..186I} and realistic three-dimensional radiative
magnetohydrodynamic simulations \citep{Rempel+etal2009Sci...325..171R, Rempel+etal2009ApJ...691..640R, Kitiashvili+etal2009ApJ...700L.178K} show that the thermal magnetoconvection is responsible for the observed Evershed effect and the filamentary intensity structures in the penumbrae. The bright penumbral grains/fibrils are locations of up-flowing hot gas associated with more vertical magnetic fields \citep[e.g.,][]{Langhans+etal2005A&A...436.1087L, rimmele+marino2006, Wang+Deng+Liu2012ApJ...748...76W}. The dark penumbral filaments are locations corresponding to cooled horizontal Evershed flows guided by more horizontal magnetic fields \citep[e.g.,][]{stanchfield+thomas+lites1997, Rempel2011ApJ...729....5R}. In the chromosphere, the filamentary penumbral structures are still present and extend out to a larger area called the superpenumbra. Inverse (i.e., inward) Evershed flow is observed in the chromospheric penumbra and superpenumbra \citep[][and references therein]{Solanki2003A&ARv..11..153S}. It is found that the penumbral filamentary structures seen at sufficiently different heights (for example, the formation heights of a strong photospheric line core and the continuum, which are a few hundreds of kilometers apart) are poorly correlated \citep{Wiehr+Stellmacher1989A&A...225..528W}, implying limited vertical extension for those features.

Besides the aforementioned long lasting and relatively stable features (usually lasting more than 20 minutes), small-scale and short lived dynamic transients are also observed in the penumbral atmosphere. Using high-resolution chromospheric \CaIIH\ filtergrams obtained by the Solar Optical Telescope \citep[SOT;][]{Tsuneta+etal2008SoPh..249..167T} onboard Hinode, \citet{Katsukawa+etal2007Sci...318.1594K} reported ``penumbral microjets'' that are small-scale jet-like brightening features (width $\sim$400 km, length 1000--4000 km, duration $<$ 1 min) ubiquitously and constantly present above sunspot penumbrae. They are found to emanate from locations near the bright penumbral grains in between dark penumbral filaments. The authors argued that the penumbral microjets might be generated by magnetic reconnection due to the drifting motion of penumbral grains in the uncombed magnetic configuration. \citet{Reardon+etal2013ApJ...779..143R} studied the spectral properties of penumbral transient brightenings in the upper photosphere and chromosphere using \CaII\ 854.21~nm spectroscopy obtained by the Interferometric Bidimensional Spectrometer \citep[IBIS;][]{Cavallini2006SoPh..236..415C}. They found that penumbral microjets show double-humped spectral line profiles (i.e., emission in both wings) that are similar to those of Ellerman bombs \citep{ellerman17}. Moreover, they identified other types of transient brightenings that show significantly different spectral characteristics, such as acoustic shocks, umbral flashes, and ``chromospheric modification". They concluded that the apparent highly dynamic brightenings above sunspots can be produced by different physical mechanisms.

The gas in the transition region (TR) above sunspots is even more complex and dynamic \citep{Solanki2003A&ARv..11..153S}. The recently launched Interface Region Imaging Spectrograph \citep[IRIS;][]{DePontieu.IRIS.2014SoPh..289.2733D} mission provides unprecedented observations of the solar TR and chromosphere with high cadence (several seconds), high spatial (0\arcsec.167~pixel$^{-1}$) and spectral resolution. Using IRIS 1400~\AA\ and 1330~\AA\ slit-jaw images (SJIs), \citet{Tian+etal.BD.2014ApJ...790L..29T} revealed numerous short-lived subarcsecond bright dots (BDs) in the TR above every sunspot they investigated (located mainly in the penumbrae). These TR penumbral BDs often have intensities a few times stronger than their surrounding background in the 1400~\AA\ SJIs. They generally appear slightly elongated along the sunspot radial direction with typical sizes of a few hundreds of kilometers. Despite some long-lasting (several minutes or even longer) strong dots, most of the BDs last no more than a minute. When a BD occurs, the TR line profiles (e.g., \SiIV\ 1402.77 \AA\ and \CII\ 1334.53 \AA) are strongly enhanced and broadened, suggesting strong emission in the TR. Some strong (i.e., very bright, relatively large and long-lasting) dots can also be seen in the 304 \AA, 171 \AA, 193 \AA, 211 \AA, and 131 \AA\ passbands of Atmospheric Imaging Assembly \citep[AIA;][]{Lemen+etal2012SoPh..275...17L} onboard Solar Dynamics Observatory \citep[SDO;][]{Pesnell.SDO.2012SoPh..275....3P}. They appear at the TR footpoints of coronal loops and are likely related to small-scale energy release events there. The thermal energy of the BDs is estimated to be in the order of 10$^{22}$--10$^{23}$ erg, which is in the scope of nanoflares \citep{ParkerNanoFlare1988ApJ...330..474P}. Using High-resolution Coronal imager (Hi-C) 193 \AA\ images, \citet{Alpert+etal2016ApJ...822...35A} also observed BDs in a sunspot penumbra and performed a similar analysis to characterize their physical properties. Hi-C penumbral BDs are on average slower, dimmer, larger in size and longer lived than IRIS penumbral BDs. Analysis of light curves of different AIA passbands suggests that the temperatures of most Hi-C penumbral BDs are likely in the TR range. It is, however, not clear yet whether these TR penumbral BDs have any signatures in the lower atmosphere and how are they formed. \citet{Tian+etal.BD.2014ApJ...790L..29T} proposed two possible generation mechanisms for them: 1. small-scale magnetic reconnection in the TR and chromosphere involving the uncombed penumbral magnetic fields; 2. associated with falling plasma.

To fully understand the origin of the TR penumbral subarcsecond BDs and their relationship with the aforementioned lower-atmospheric penumbral features, co-temporal observations in the chromosphere and photosphere with high spatial and temporal resolution are necessary. This paper presents a joint study of these BDs using a coordinated observation by IRIS and the 1.6~m New Solar Telescope \citep[NST;][]{GoodeNST2010AN....331..620G} at the Big Bear Solar Observatory (BBSO). The \ha\ imaging spectroscopy and the TiO images from BBSO/NST provide important information in the chromosphere and photosphere, respectively, which enables us to investigate whether and to what extent a correspondence of the TR penumbral BDs exists in the lower atmosphere.

\section{OBSERVATIONS}\label{sec:observation}

\begin{figure*}[t]
\epsscale{1.15}
\plotone{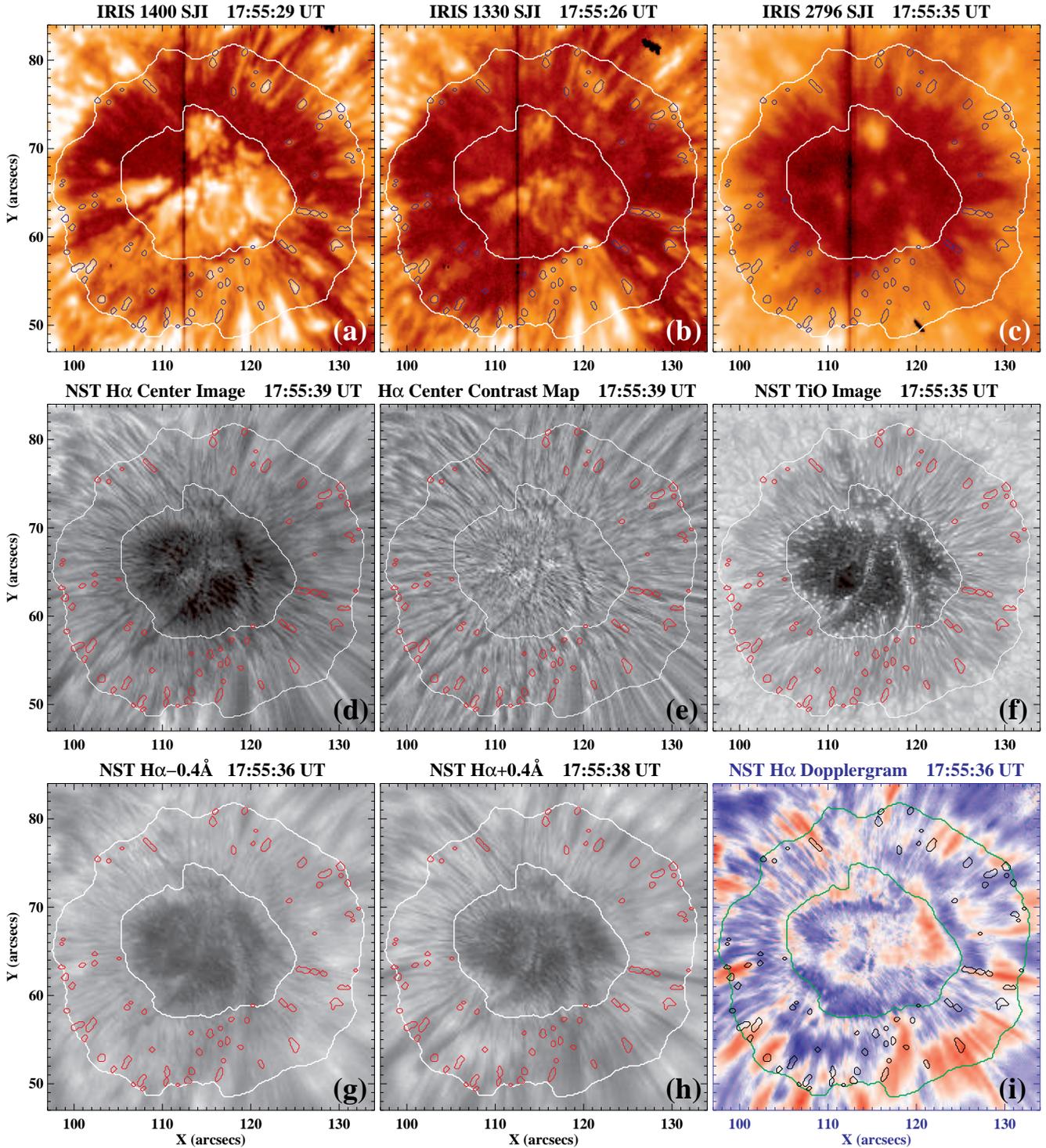}
\caption{Joint IRIS/BBSO observation on 2013/09/02 for the main sunspot in NOAA AR 11836. (a)--(c): The IRIS \SiIV\ 1400, \CII\ 1330, and \MgII\ k 2796~\AA\ SJIs. (d) The BBSO/NST \ha\ narrow-band image at the line center. (e) The contrast map (defined by Equation~1) of the \ha\ line center image. (f) The BBSO/NST photospheric TiO broad-band image. (g)--(h): The \ha\ narrow-band images at the blue and red wing. (i) \ha\ Dopplergram generated by subtracting the H$\alpha+0.4$~\AA\ image from the H$\alpha-0.4$~\AA\ image. The colors scale from -10 (blue) to +10 (red) \kms. The white contours mark the umbral and penumbral boundaries that are obtained from a smoothed TiO image. The TR penumbral BDs that are identified using the IRIS 1400~\AA\ SJI are plotted as small contours within the penumbral area on each panel.\\
{\vskip -2mm (A color version of this figure and an animation are available in the online journal.)} \vskip 2mm} 
\label{FIG:co-images}
\end{figure*}

On 2013 September 2, the IRIS and BBSO/NST jointly observed the main sunspot in NOAA AR 11836, which was close to the disc center ($\mu=cos\theta=0.97$) as shown in Figure~\ref{FIG:co-images}. The obtained data have been used to study the sunspot oscillation in the TR and chromosphere by \citet{Tian+etal.Osci.2014ApJ...786..137T}, \citet{Madsen+etal2015ApJ...800..129M} and \citet{Yurchyshyn+etal2015ApJ...798..136Y}. In this paper, a continuous 37-minute data set (17:21--17:58 UT) during a good and stable seeing period at the NST is selected from the full IRIS data set (16:39--17:58 UT) to study the TR penumbral BDs and their possible signatures in the chromosphere and photosphere. 

IRIS observed in sit-and-stare mode with a cadence of 3~s for the FUV and NUV spectra covering a medium line list. The SJIs have a cadence of 12~s in each filter of the three observed passbands: \SiIV\ 1400, \CII\ 1330, and \MgII\ k 2796~\AA. These SJIs are dominated by emission from, respectively, the \SiIV\ 1393.76/1402.77~\AA\ lines that are formed in the middle TR ($10^{4.9}$ K), the \CII\ 1334.53/1335.71~\AA\ lines that are formed in the lower TR ($10^{4.4}$ K), and the \MgII~k 2796.35~\AA~line that is formed in the upper chromosphere ($10^{4.0}$ K). The image scale and field-of-view (FOV) of the SJIs are 0\arcsec.167~pixel$^{-1}$ and $60^{\prime\prime} \times 61^{\prime\prime}$, respectively. The spatial resolution is 0\arcsec.33 for the FUV (\SiIV\ and \CII) channels and 0\arcsec.4 for the NUV (\MgII\ k) channel. The calibrated level 2 data were used in this study.

The BBSO/NST obtained observations with real time seeing correction by using the high-order adaptive optics of 308 subapertures \citep[AO-308;][]{Shumko+etal2014SPIE.9148E..35S} and with post-facto speckle image reconstruction \citep{Woger+Luhe2007ApOpt..46.8015W}. Time series of photospheric 7057~\AA\ broadband (10~\AA) TiO images (as a proxy for continuum) were obtained by the Broadband Filter Imagers (BFIs) with an image scale of 0\arcsec.0375~pixel$^{-1}$ and a time cadence of 15~s after speckle reconstruction. The chromospheric \ha\ imaging spectroscopy was obtained by the Visible Imaging Spectrometer (VIS), which combines a 5~\AA\ interference filter and a tunable Fabry-P\'erot etalon to acquire narrow-band (0.07~\AA) images at five wavelength positions ($\mp 0.8$~\AA, $\mp 0.4$~\AA, and 0.0~\AA) across the \ha\ line center (6562.8~\AA). At each wavelength position, 25 images were taken as a burst for post-facto speckle reconstruction. The time lapse between two sequential wavelength positions (e.g., $-0.4$~\AA\  and $+0.4$~\AA) is about 2~s. The cadence of performing each five-point scan was set to 23~s. The image scale and FOV of the \ha\ narrow-band images are 0\arcsec.029~pixel$^{-1}$ and $70^{\prime\prime} \times 70^{\prime\prime}$, respectively. \ha\ Doppler maps are generated by subtracting the H$\alpha+0.4$~\AA\ image from the corresponding H$\alpha-0.4$~\AA\ image. Note that the diffraction limited resolution of the 1.6~m NST is 0\arcsec.111 and 0\arcsec.103 at TiO and \ha\ wavelengths, respectively, if we use 1.22$\lambda/D$ as the measure of the diffraction limit \citep{Cao+etal2010SPIE.7735E..5VC}. Therefore the images were oversampled a little bit.

\section{DATA ANALYSIS}\label{sec:reduction}

The IRIS and NST data sets were carefully co-aligned by manually registering prominent features in the umbra and brightening events outside the sunspot. The error of co-alignment is estimated to be no more than $0.5^{\prime\prime}$ \citep{Yurchyshyn+etal2015ApJ...798..136Y}.

As can be seen from Figure~\ref{FIG:co-images} (a) and (b), in the TR, the quiet penumbral area is relatively darker than the umbral and outside sunspot regions. Bright features are present almost everywhere. Those inside penumbrae are more compact, isolated and ephemeral. In this study, we only focus on the subarcsecond and short-lived BDs within the penumbral area studied by \citet{Tian+etal.BD.2014ApJ...790L..29T}. 

We use IRIS \SiIV\ 1400~\AA\ SJIs to identify TR penumbral BDs by following similar methods used in \citet{Tian+etal.BD.2014ApJ...790L..29T}. Each image is first boxcar smoothed using an averaging kernel of $9 \times 9$ pixels then subtracted from the original image to sharpen the boundaries of BDs. The BDs can be outlined using intensity contours of the sharpened images. The level of the contour is set to 15 (in units of DNs, here, DNs refers to data numbers), which is equivalent to a contrast threshold of about 110\%. We exclude large-scale bright loops or patches having sizes larger than $1^{\prime\prime}$. We also exclude very small events having only one pixel that are presumably caused by spikes or noise. Each BD identified from a frame is examined if it has any spatial overlap with a BD identified from its adjacent frames. If no overlap is found, we assume that the BD only appears in one frame. If an overlap is found, the dots detected in consecutive frames will be counted as one BD event. The number of frames (could be one) during which a BD exists gives its lifetime. A total of 2692 penumbral BDs were automatically detected from the 37-minute time series containing 187 frames of the 1400~\AA\ SJIs. 


\begin{figure}[t]
\epsscale{1.05}
\plotone{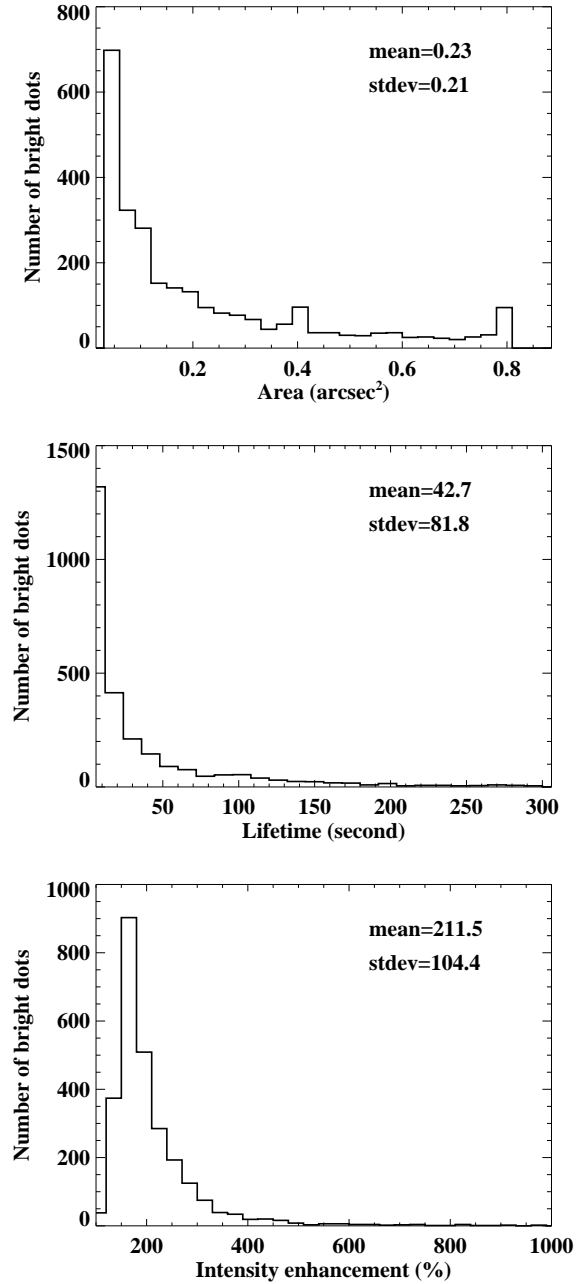}
\caption{Statistical properties of the 2692 penumbral BDs detected from IRIS \SiIV\ 1400~\AA\ SJIs. The area and intensity contrast were measured at the peak (i.e., brightest) time of each BD.\\}
\label{FIG:bd_histo}
\end{figure}

Figure~\ref{FIG:bd_histo} shows the distributions of the area, lifetime, and intensity enhancement of the 2692 BDs measured based on IRIS 1400 SJIs. The mean of the area (0.23 arcsec$^2$) gives an equivalent size of 0\arcsec.54 (i.e., 396~km) if assuming the BDs have a round shape. The intensity enhancement with respect to surrounding background is obtained from the contrast map that is defined as:
\begin{equation}
Contrast \ Map = \frac{image-smooth(image)}{smooth(image)} \ , 
\end{equation}
where smooth(image) means the image is boxcar smoothed over an area of $15 \times 15$ pixels, which gives a good approximation for the background intensity.

These statistical properties are very similar to those presented in \citet{Tian+etal.BD.2014ApJ...790L..29T} that were based on a sample of 176 manually selected BDs, thus validating our automatic detection of these events. The automatic detection may include more weak events comparing to human selection, therefore a little bit lower value of the mean intensity enhancement is obtained for our BD sample.  

We then spatiotemporally map the detected 1400~\AA\ BDs onto co-aligned other wavelengths image sequences observed by IRIS (1330~\AA\ and 2796~\AA) and NST (\ha\ and TiO). These wavelengths diagnose progressively lower atmospheres \citep[e.g.,][]{Leenaarts+etal2013ApJ...772...90L, DePontieu.IRIS.2014SoPh..289.2733D}.

\section{RESULTS}\label{sec:results}

\subsection{Preferred Locations}\label{sec:location}

\citet{Tian+etal.BD.2014ApJ...790L..29T} observed that the BDs seem to be located in or at the edge of bright filamentary structures in the TR. The data shown in this paper confirm this point (see Figure~\ref{FIG:co-images} (a), (b) and accompanied online movie m1.mpg). \citet{Tian+etal.BD.2014ApJ...790L..29T} also found that some dots appear to be associated with bright penumbral filamentary structures in the photosphere shown by the HMI continuum images. A solid conclusion with certain statistical significance of the preferred locations of the TR penumbral BDs in the chromosphere and photosphere requires a statistical study of a large enough sample, which is carried out in this paper.

\begin{figure}[t]
\epsscale{1.15}
\plotone{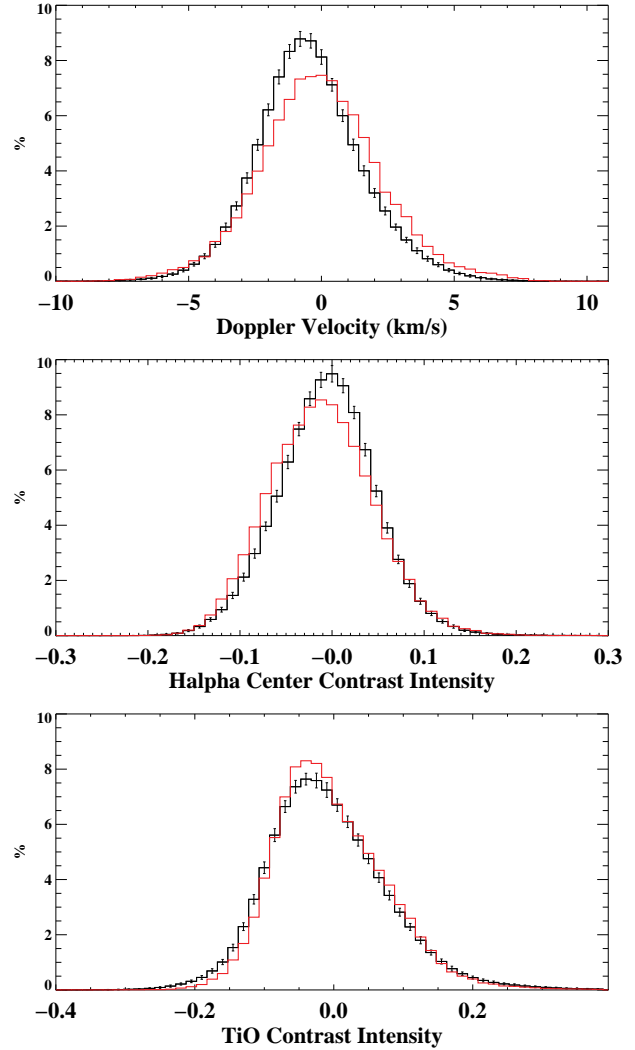}
\caption{Red lines: distributions of locations of TR penumbral BDs in \ha\ Dopplergrams (a), \ha\ line center contrast maps (b), and TiO contrast maps (c). Black lines: distributions of randomly selected locations within the penumbra with 3$\sigma$ error bars plotted. Negative/positive values represent blue/red shifts (a) or darker/brighter features (b and c). \\
{\vskip -2mm (A color version of this figure is available in the online journal.)} \vskip 2mm} 
\label{FIG:location}
\end{figure}

Figure~\ref{FIG:location} shows the distributions of locations of the TR penumbral BDs in \ha\ and TiO observations, comparing to those of randomly selected locations in the penumbra. The randomly selected locations have the same pixel number as that of BDs in each frame and are uniformly distributed across the penumbra. In order to discriminate bright and dark features regardless of radial position, we use contrast maps (an example is shown in Figure~\ref{FIG:co-images} (e)), which eliminate the dependence of intensity on the radial distance from the spot center. Comparing to the black curves (i.e., random locations), the red curves (i.e., BDs locations) shift significantly (with a statistical significance considerably larger than 3$\sigma$) toward the redshift regions in \ha\ Dopplergram (panel a) and toward the darker features in \ha\ line center contrast maps (panel b). Considering that the sunspot was close to the disk center, the redshifts mostly represent downflows. Darker \ha\ penumbral features generally indicate denser and/or cooler fibrils that presumably trace magnetic flux tubes \citep{delaCruz+Socas-Navarro2011A&A...527L...8D, Schad+etal2013ApJ...768..111S, Leenaarts+etal2015ApJ...802..136L}. These results suggest that TR penumbral BDs tend to be associated more with downflowing and dense/cool magnetic flux tubes in the chromosphere. 

\begin{figure}[t]
\epsscale{1.15}
\plotone{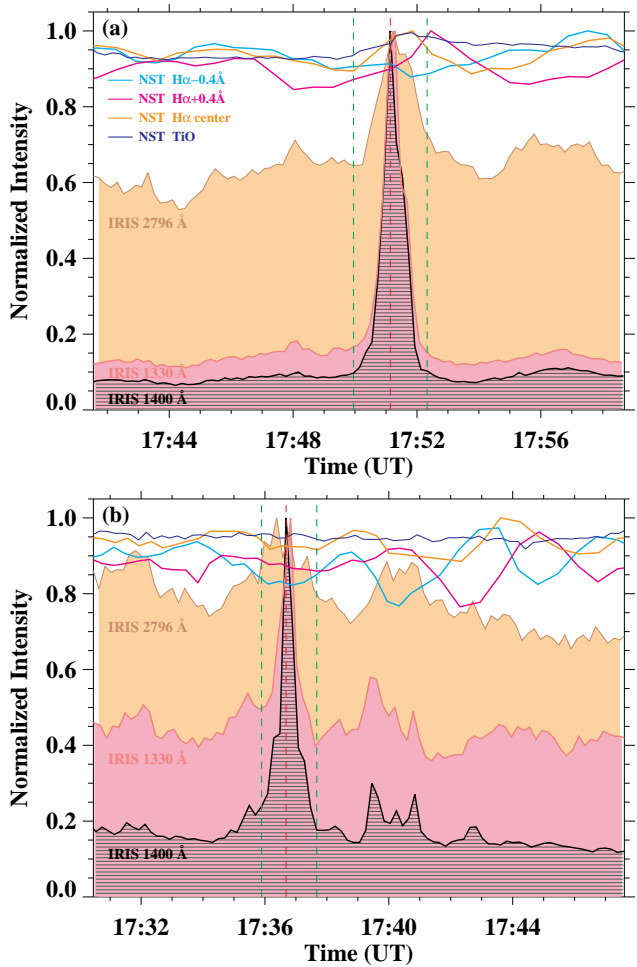}
\caption{The multi-wavelength light curves (i.e., temporal evolution of normalized intensities averaged over the area of a BD) for two BDs (panel a and b). IRIS light curves are filled for a clearer illustration. The red vertical dashed line indicates the peak time of the BD in IRIS 1400~\AA\ observation. The green vertical dashed lines outline the time window of the BD, over which linear temporal correlations between IRIS 1400~\AA\ and other wavelengths are calculated and presented in Figure~\ref{FIG:cc}.\\
{\vskip -2mm (A color version of this figure is available in the online journal.)} \vskip 2mm} 
\label{FIG:cases}
\end{figure}

\begin{figure*}[t]
\epsscale{1.15}
\plotone{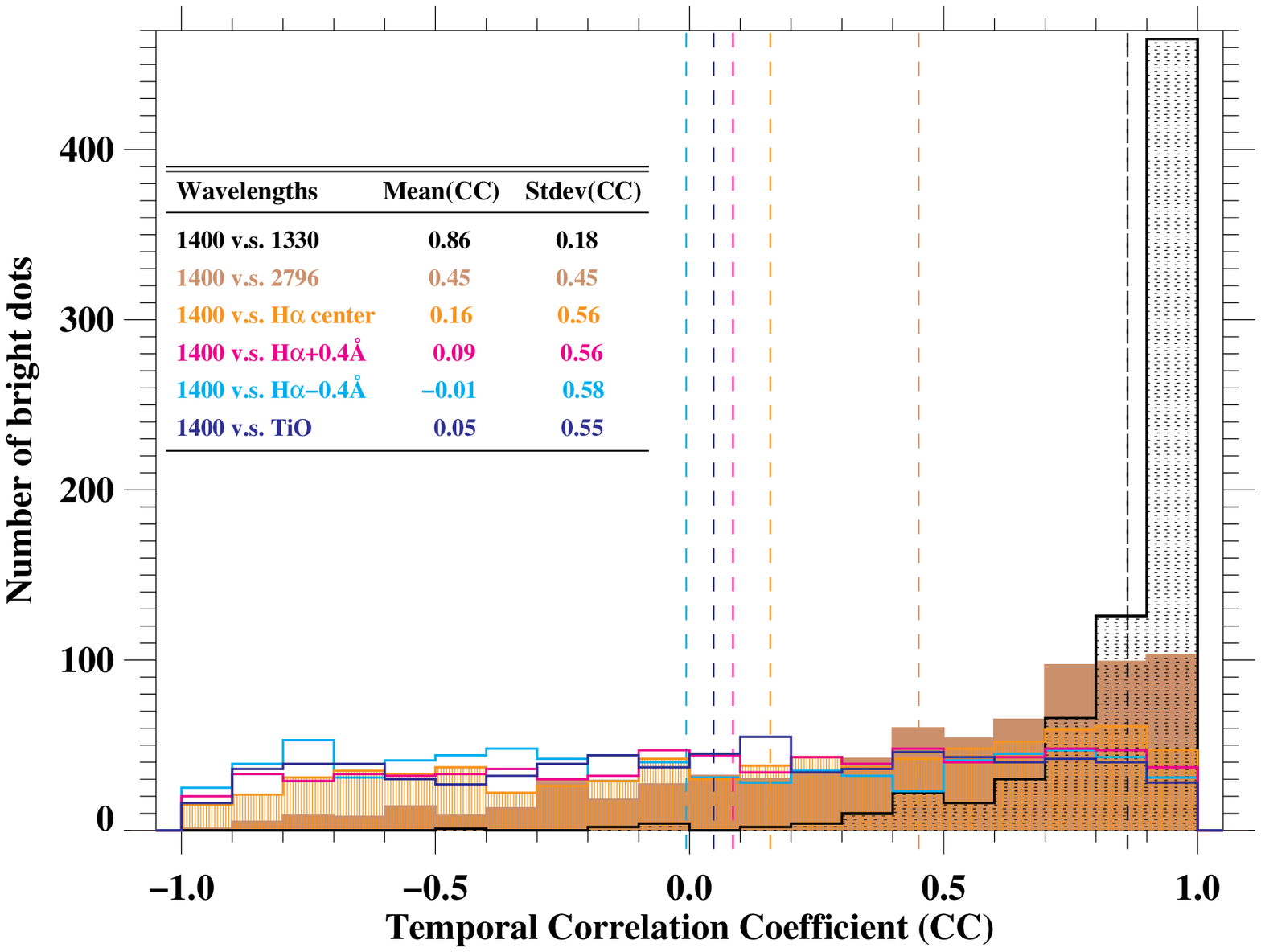}
\caption{The distributions of temporal correlation coefficients (CC) between IRIS 1400~\AA\ and other wavelengths (represented by different colors) calculated during the course of 748 BDs. Their mean and standard deviation (stdev) values are tabled on the figure with corresponding colors. The mean values for each histogram are also indicated by the vertical dashed lines of corresponding colors. \\
{\vskip -2mm (A color version of this figure is available in the online journal.)} \vskip 2mm} 
\label{FIG:cc}
\end{figure*}

The distribution of BDs locations shifts only slightly (but still exceeds 3$\sigma$ in most bins) toward brighter features in TiO contrast maps (panel c of Figure~\ref{FIG:location}), suggesting a weak association with bright penumbral structures in the photosphere. This is consistent with \citet{Tian+etal.BD.2014ApJ...790L..29T}'s observation. In the photosphere, bright penumbral structures generally have a more vertical magnetic orientation than the dark penumbral filaments (e.g., see Figure~1 in \citet{Wang+Deng+Liu2012ApJ...748...76W} and Figure~13 in \citet{Langhans+etal2005A&A...436.1087L}). 

To summarize, the above results and analyses suggest that TR penumbral BDs prefer to locate along more vertical magnetic loops with downflowing dense and cool materials, which is compatible with the falling plasma scenario. Indeed, visual inspection of the movie (m1.mpg) reveals that some BDs are clearly associated with falling materials shown in IRIS 1400 and 1330~\AA\ time series and with cotemporal redshifts shown in \ha\ Dopplergrams. The falling plasmas could be due to siphon flows, free fall, coronal rain, or reconnection outflows generated at TR or coronal heights. 

\subsection{Temporal Response in Lower Atmospheres}\label{sec:temporal}

Studying and comparing the temporal responses of the BDs at different atmospheric heights/wavelengths can provide important clues on their origin and formation mechanism. Based on several cases whose spectra were analyzed, \citet{Tian+etal.BD.2014ApJ...790L..29T} found decreased emission response from the TR down to the upper chromosphere. BDs usually cause strong emission in the middle-TR line of the \SiIV\ 1402.77~\AA\ whose intensity can be enhanced by one to two orders of magnitude. In the lower TR,  the intensity of the \CII\ 1334.53~\AA\ line is only enhanced by a few factors. The upper chromospheric \MgII~K 2796.35~\AA\ line shows a much weaker intensity enhancement, suggesting insignificant chromospheric emission.

In this paper, we systematically study the light curves of the 2692 BDs using multi-wavelength observations that cover the middle TR (1400~\AA), lower TR (1330~\AA), upper chromosphere (2796~\AA), middle and lower chromosphere (\ha\ center and wings), as well as the photosphere (TiO). Figure~\ref{FIG:cases} shows the temporal evolution of two BDs at these wavelengths. When a BD occurs, the intensity in 1400~\AA\ SJI can increase by up to one order of magnitude. The light curves of 1330~\AA\ usually highly correlate with those of 1400~\AA, but having a smaller intensity enhancement during the BD. The brightening response in 2796~\AA\ is much weaker, with an intensity elevated to no more than twice. These results using IRIS SJI observations are consistent with \citet{Tian+etal.BD.2014ApJ...790L..29T}'s result using spectra observation. The response of the BDs in \ha\ observations are inconsistent. The \ha\ intensities can either increase (panel a) or decrease (panel b) for different BDs. Generally, the \ha\ responses are unnoticeable because the intensity variations are not well above the noise level. Therefore, there is no consistent and prominent response in the middle and lower chromosphere. Similarly, no consistent and noticeable response is found in the photosphere.

To quantify the temporal correspondence of BDs at different wavelengths (heights), we calculate the temporal correlation between IRIS 1400~\AA\ and other wavelengths light curves for those relatively longer-lived BDs that last at least 4 frames of 1400~\AA\ (748 out of 2692). The temporal correlation is calculated within a time window around the lifetime of each BD. Such a time window is indicated by the green vertical dashed lines in Figure~\ref{FIG:cases}.  The time window covers the lifetime of each BD and also includes two more frames before the BD is first detected and two more frames after it is last detected. Figure~\ref{FIG:cc} summarizes the result in terms of histograms of the temporal Correlation Coefficients (CC). The CCs between IRIS 1400~\AA\ and 1330~\AA\ (shown by the black histogram) are concentrated in the $> 0.7$ region with a mean of 0.86, meaning 1330~\AA\ and 1400~\AA\ light curves are highly correlated for most BDs. The distribution of CCs between IRIS 1400~\AA\ and 2796~\AA\ (the brown solid filled histogram) is more spread toward the lower even negative values with a mean CC of 0.45, suggesting a much weaker temporal correspondence in 2796~\AA\ for the TR BDs. The distributions of CCs between  IRIS 1400~\AA\ and \ha\ as well as TiO light curves are almost flatly spread all over the entire range resulting in mean values of nearly zero, except for \ha\ center (the orange histogram) which has a little preference in the positive end and a mean CC of 0.16. These results suggest that TR penumbral BDs have no response at all in the lower chromosphere and photosphere, maybe a little brightening signature in the middle chromosphere.

If the BDs are generated by small-scale impulsive magnetic reconnection, the above results suggest that such reconnection must occur sufficiently high in the atmosphere, i.e., in the middle TR or above, so that the brightening response dramatically decreases with depth and there is no noticeable energy release in the chromosphere.

\section{SUMMARY AND DISCUSSIONS}\label{sec:summary}

We have presented a statistical study of TR penumbral subarcsecond BDs using multi-wavelength observations from IRIS and NST. A total of 2692 BDs are automatically identified from a 37-minute time series of IRIS 1400~\AA\ SJIs. They have a mean size of $\sim$400~km and a mean lifetime of $\sim$43~s. These statistical properties are very similar to those presented in \citet{Tian+etal.BD.2014ApJ...790L..29T}. The locations of TR penumbral BDs tend to be associated more with redshifted (i.e., downflowing) and darker features in the chromosphere, and weakly associated with brighter features in the photosphere. Analyses of multi-wavelength temporal light curves of the BDs show that the brightening response dramatically decreases with depth. Only weak brightening response is found in the upper chromosphere. There is no noticeable and consistent response below the middle chromosphere. 

Our results exclude the possibility that the TR penumbral BDs are directly generated by magnetic reconnection taking place in the chromosphere, because no convincing energy release signature was found in the chromosphere. It is still possible that the small-scale magnetic reconnection occurs sufficiently high in the atmosphere, i.e., in the middle TR or above, which causes localized impulsive energy release. The falling plasma scenario is compatible with our observational results, therefore it is also very likely. 

Subarcsecond heating/brightening events are recently observed in the TR or corona in different places by IRIS and Hi-C. For example, using IRIS observations, \citet{Kleint+etal2014ApJ...789L..42K} found small-scale heating events associated with supersonic downflows in the TR above sunspot umbrae and explained them with falling plasma mechanism. \citet{Skogsrud+etal2016ApJ...817..124S} studied bright roundish small patches, so-called bright ``grains", which ubiquitously appear in IRIS 1400 channel in active region plages. The bright grains have sizes of 0\arcsec.5 -- 1\arcsec.7 and lifetimes of a few minutes. Many of them show good correspondence with the behavior of shock-driven ``dynamic fibrils" observed in \ha. The authors thus suggest that the bright grains are the result of chromospheric shocks impacting the TR. Using Hi-C observation, \citet{Regnier+etal2014ApJ...784..134R} reported sparkling extreme-ultraviolet (EUV) BDs at the edge of an active region. These EUV BDs have similar sizes and lifetimes as the TR penumbral BDs, although they occur at a different place. They are interpreted as small-scale impulsive heating events at the base of coronal loops.

Combining observations from Swedish 1-m Solar Telescope (SST) in \CaII\ lines and IRIS for a near limb sunspot, \citet{Vissers+etal2015ApJ...811L..33V} found evidence for a TR brightening response to chromospheric penumbral microjets along the jet extent. In contrast to our analysis, they used an opposite approach. They first identify chromospheric penumbral bright jets using SST \CaII\ data, then seek TR response signatures using IRIS data. Considering the similar sizes and lifetimes between the penumbral microjets and the TR penumbral BDs, the authors speculate that many moving TR BDs are the tops of these jets. We cannot exclude that some TR BDs might be related to the chromospheric penumbral bright jets, considering the weak emission response in 2796~\AA\ and a small fraction showing positive correlation with \ha\ center light curves. However, they could not be the majority. \citet{Tian+etal.BD.2014ApJ...790L..29T} reported that about 32\% of BDs move inward and only 13\% move outward. The jet top scenario maybe more suitable for those outward moving BDs which only account for a small fraction. The inward moving BDs seem favor more falling motion, rather than jetting motion. Using Hinode/SOT, Hi-C, and SDO/AIA observations of a sunspot, \citet{Tiwari+etal2016ApJ...816...92T} also examined whether chromospheric penumbral microjets have signatures in the TR and corona. They found that most penumbral microjets hardly have any discernible signature in most AIA passbands except for some exceptionally stronger/larger jets occurred in the tails of some penumbral filaments where opposite-polarity magnetic fields are seen. These tail penumbral jets do have brightening signatures in the TR.


Future development of high-resolution magnetic field measurements in the chromosphere \citep[e.g.,][]{delaCruz+Socas-Navarro2011A&A...527L...8D, Rouppe+delaCruz2013ApJ...776...56R, Schad+etal2013ApJ...768..111S} will allow us to directly investigate the relation of BDs to the magnetic field topology of the lower solar atmosphere.

\acknowledgments
N.D., C.L., Y.X., and H.W. were partially supported by NASA grants NNX13AF76G, NNX13AG13G, and NNX14AC12G, and by NSF grants AGS 1348513, AGS 1408703, and AGS 1539791. V.Y. acknowledges support from AFOSR FA9550-12-0066 and NSF AGS-1146896 grants and the Korea Astronomy and Space Science Institute. H.T. was supported by the Recruitment Program of Global Experts of China and the NSFC grant 41574166. We thank the referee for helpful comments that improve the paper. IRIS is a NASA small explorer mission developed and operated by LMSAL with mission operations executed at NASA Ames Research center and major contributions to downlink communications funded by ESA and the Norwegian Space Centre. BBSO operation is supported by NJIT, US NSF AGS-1250818 and NASA NNX13AG14G, and NST operation is partly supported by the Korea Astronomy and Space Science Institute and Seoul National University, and by strategic priority research program of CAS with Grant No. XDB09000000.


\begin{thebibliography}{38}
\expandafter\ifx\csname natexlab\endcsname\relax\def\natexlab#1{#1}\fi

\bibitem[{{Alpert} {et~al.}(2016){Alpert}, {Tiwari}, {Moore}, {Winebarger}, \&
  {Savage}}]{Alpert+etal2016ApJ...822...35A}
{Alpert}, S.~E., {Tiwari}, S.~K., {Moore}, R.~L., {Winebarger}, A.~R., \&
  {Savage}, S.~L. 2016, \apj, 822, 35

\bibitem[{{Bellot Rubio} {et~al.}(2004){Bellot Rubio}, {Balthasar}, \&
  {Collados}}]{BellotRubio+etal2004A&A...427..319B}
{Bellot Rubio}, L.~R., {Balthasar}, H., \& {Collados}, M. 2004, \aap, 427, 319

\bibitem[{{Borrero} \& {Ichimoto}(2011)}]{Borrero+Ichimoto2011LRSP....8....4B}
{Borrero}, J.~M., \& {Ichimoto}, K. 2011, Living Reviews in Solar Physics, 8, 4

\bibitem[{{Cao} {et~al.}(2010){Cao}, {Gorceix}, {Coulter}, {W{\"o}ger}, {Ahn},
  {Shumko}, {Varsik}, {Coulter}, \& {Goode}}]{Cao+etal2010SPIE.7735E..5VC}
{Cao}, W., {Gorceix}, N., {Coulter}, R., {W{\"o}ger}, F., {Ahn}, K., {Shumko},
  S., {Varsik}, J., {Coulter}, A., \& {Goode}, P.~R. 2010, in Society of
  Photo-Optical Instrumentation Engineers (SPIE) Conference Series, Vol. 7735,
  Ground-based and Airborne Instrumentation for Astronomy III, 77355V

\bibitem[{{Cavallini}(2006)}]{Cavallini2006SoPh..236..415C}
{Cavallini}, F. 2006, \solphys, 236, 415

\bibitem[{{de la Cruz Rodr{\'{\i}}guez} \&
  {Socas-Navarro}(2011)}]{delaCruz+Socas-Navarro2011A&A...527L...8D}
{de la Cruz Rodr{\'{\i}}guez}, J., \& {Socas-Navarro}, H. 2011, \aap, 527, L8

\bibitem[{{De Pontieu} {et~al.}(2014){De Pontieu}, {Title}, {Lemen},
  {et~al.}}]{DePontieu.IRIS.2014SoPh..289.2733D}
{De Pontieu}, B., {Title}, A.~M., {Lemen}, J.~R., {et~al.} 2014, \solphys, 289,
  2733

\bibitem[{{Deng} {et~al.}(2010){Deng}, {Prasad Choudhary}, \&
  {Balasubramaniam}}]{Deng+etal2010ApJ...719..385D}
{Deng}, N., {Prasad Choudhary}, D., \& {Balasubramaniam}, K.~S. 2010, \apj,
  719, 385

\bibitem[{{Ellerman}(1917)}]{ellerman17}
{Ellerman}, F. 1917, \apj, 46, 298

\bibitem[{{Goode} {et~al.}(2010){Goode}, {Coulter}, {Gorceix}, {Yurchyshyn}, \&
  {Cao}}]{GoodeNST2010AN....331..620G}
{Goode}, P.~R., {Coulter}, R., {Gorceix}, N., {Yurchyshyn}, V., \& {Cao}, W.
  2010, Astronomische Nachrichten, 331, 620

\bibitem[{{Ichimoto}(2010)}]{Ichimoto2010mcia.conf..186I}
{Ichimoto}, K. 2010, in Magnetic Coupling between the Interior and Atmosphere
  of the Sun, ed. {S.~S.~Hasan \& R.~J.~Rutten}, 186--192

\bibitem[{{Ichimoto} {et~al.}(2007){Ichimoto}, {Shine}, {Lites}, {Kubo},
  {Shimizu}, {Suematsu}, {Tsuneta}, {Katsukawa}, {Tarbell}, {Title}, {Nagata},
  {Yokoyama}, \& {Shimojo}}]{Ichimoto+etal2007PASJ...59S.593I}
{Ichimoto}, K., {Shine}, R.~A., {Lites}, B., {Kubo}, M., {Shimizu}, T.,
  {Suematsu}, Y., {Tsuneta}, S., {Katsukawa}, Y., {Tarbell}, T.~D., {Title},
  A.~M., {Nagata}, S., {Yokoyama}, T., \& {Shimojo}, M. 2007, \pasj, 59, 593

\bibitem[{{Katsukawa} {et~al.}(2007){Katsukawa}, {Berger}, {Ichimoto}, {Lites},
  {Nagata}, {Shimizu}, {Shine}, {Suematsu}, {Tarbell}, {Title}, \&
  {Tsuneta}}]{Katsukawa+etal2007Sci...318.1594K}
{Katsukawa}, Y., {Berger}, T.~E., {Ichimoto}, K., {Lites}, B.~W., {Nagata}, S.,
  {Shimizu}, T., {Shine}, R.~A., {Suematsu}, Y., {Tarbell}, T.~D., {Title},
  A.~M., \& {Tsuneta}, S. 2007, Science, 318, 1594

\bibitem[{{Kitiashvili} {et~al.}(2009){Kitiashvili}, {Kosovichev}, {Wray}, \&
  {Mansour}}]{Kitiashvili+etal2009ApJ...700L.178K}
{Kitiashvili}, I.~N., {Kosovichev}, A.~G., {Wray}, A.~A., \& {Mansour}, N.~N.
  2009, \apjl, 700, L178

\bibitem[{{Kleint} {et~al.}(2014){Kleint}, {Antolin}, {Tian}, {Judge}, {Testa},
  {De Pontieu}, {Mart{\'{\i}}nez-Sykora}, {Reeves}, {Wuelser}, {McKillop},
  {Saar}, {Carlsson}, {Boerner}, {Hurlburt}, {Lemen}, {Tarbell}, {Title},
  {Golub}, {Hansteen}, {Jaeggli}, \&
  {Kankelborg}}]{Kleint+etal2014ApJ...789L..42K}
{Kleint}, L., {Antolin}, P., {Tian}, H., {Judge}, P., {Testa}, P., {De
  Pontieu}, B., {Mart{\'{\i}}nez-Sykora}, J., {Reeves}, K.~K., {Wuelser},
  J.~P., {McKillop}, S., {Saar}, S., {Carlsson}, M., {Boerner}, P., {Hurlburt},
  N., {Lemen}, J., {Tarbell}, T.~D., {Title}, A., {Golub}, L., {Hansteen}, V.,
  {Jaeggli}, S., \& {Kankelborg}, C. 2014, \apjl, 789, L42

\bibitem[{{Langhans} {et~al.}(2005){Langhans}, {Scharmer}, {Kiselman},
  {L{\"o}fdahl}, \& {Berger}}]{Langhans+etal2005A&A...436.1087L}
{Langhans}, K., {Scharmer}, G.~B., {Kiselman}, D., {L{\"o}fdahl}, M.~G., \&
  {Berger}, T.~E. 2005, \aap, 436, 1087

\bibitem[{{Leenaarts} {et~al.}(2015){Leenaarts}, {Carlsson}, \& {Rouppe van der
  Voort}}]{Leenaarts+etal2015ApJ...802..136L}
{Leenaarts}, J., {Carlsson}, M., \& {Rouppe van der Voort}, L. 2015, \apj, 802,
  136

\bibitem[{{Leenaarts} {et~al.}(2013){Leenaarts}, {Pereira}, {Carlsson},
  {Uitenbroek}, \& {De Pontieu}}]{Leenaarts+etal2013ApJ...772...90L}
{Leenaarts}, J., {Pereira}, T.~M.~D., {Carlsson}, M., {Uitenbroek}, H., \& {De
  Pontieu}, B. 2013, \apj, 772, 90

\bibitem[{{Lemen} {et~al.}(2012){Lemen}, {Title}, {Akin},
  {et~al.}}]{Lemen+etal2012SoPh..275...17L}
{Lemen}, J.~R., {Title}, A.~M., {Akin}, D.~J., {et~al.} 2012, \solphys, 275, 17

\bibitem[{{Lites} {et~al.}(1993){Lites}, {Elmore}, {Seagraves}, \&
  {Skumanich}}]{Lites+etal1993ApJ...418..928L}
{Lites}, B.~W., {Elmore}, D.~F., {Seagraves}, P., \& {Skumanich}, A.~P. 1993,
  \apj, 418, 928

\bibitem[{{Madsen} {et~al.}(2015){Madsen}, {Tian}, \&
  {DeLuca}}]{Madsen+etal2015ApJ...800..129M}
{Madsen}, C.~A., {Tian}, H., \& {DeLuca}, E.~E. 2015, \apj, 800, 129

\bibitem[{{Parker}(1988)}]{ParkerNanoFlare1988ApJ...330..474P}
{Parker}, E.~N. 1988, \apj, 330, 474

\bibitem[{{Pesnell} {et~al.}(2012){Pesnell}, {Thompson}, \&
  {Chamberlin}}]{Pesnell.SDO.2012SoPh..275....3P}
{Pesnell}, W.~D., {Thompson}, B.~J., \& {Chamberlin}, P.~C. 2012, \solphys,
  275, 3

\bibitem[{{Reardon} {et~al.}(2013){Reardon}, {Tritschler}, \&
  {Katsukawa}}]{Reardon+etal2013ApJ...779..143R}
{Reardon}, K., {Tritschler}, A., \& {Katsukawa}, Y. 2013, \apj, 779, 143

\bibitem[{{R{\'e}gnier} {et~al.}(2014){R{\'e}gnier}, {Alexander}, {Walsh},
  {Winebarger}, {Cirtain}, {Golub}, {Korreck}, {Mitchell}, {Platt}, {Weber},
  {De Pontieu}, {Title}, {Kobayashi}, {Kuzin}, \&
  {DeForest}}]{Regnier+etal2014ApJ...784..134R}
{R{\'e}gnier}, S., {Alexander}, C.~E., {Walsh}, R.~W., {Winebarger}, A.~R.,
  {Cirtain}, J., {Golub}, L., {Korreck}, K.~E., {Mitchell}, N., {Platt}, S.,
  {Weber}, M., {De Pontieu}, B., {Title}, A., {Kobayashi}, K., {Kuzin}, S., \&
  {DeForest}, C.~E. 2014, \apj, 784, 134

\bibitem[{{Rempel}(2011)}]{Rempel2011ApJ...729....5R}
{Rempel}, M. 2011, \apj, 729, 5

\bibitem[{{Rempel} {et~al.}(2009{\natexlab{a}}){Rempel}, {Sch{\"u}ssler},
  {Cameron}, \& {Kn{\"o}lker}}]{Rempel+etal2009Sci...325..171R}
{Rempel}, M., {Sch{\"u}ssler}, M., {Cameron}, R.~H., \& {Kn{\"o}lker}, M.
  2009{\natexlab{a}}, Science, 325, 171

\bibitem[{{Rempel} {et~al.}(2009{\natexlab{b}}){Rempel}, {Sch{\"u}ssler}, \&
  {Kn{\"o}lker}}]{Rempel+etal2009ApJ...691..640R}
{Rempel}, M., {Sch{\"u}ssler}, M., \& {Kn{\"o}lker}, M. 2009{\natexlab{b}},
  \apj, 691, 640

\bibitem[{{Rimmele} \& {Marino}(2006)}]{rimmele+marino2006}
{Rimmele}, T., \& {Marino}, J. 2006, \apj, 646, 593

\bibitem[{{Rouppe van der Voort} \& {de la Cruz
  Rodr{\'{\i}}guez}(2013)}]{Rouppe+delaCruz2013ApJ...776...56R}
{Rouppe van der Voort}, L., \& {de la Cruz Rodr{\'{\i}}guez}, J. 2013, \apj,
  776, 56

\bibitem[{{Schad} {et~al.}(2013){Schad}, {Penn}, \&
  {Lin}}]{Schad+etal2013ApJ...768..111S}
{Schad}, T.~A., {Penn}, M.~J., \& {Lin}, H. 2013, \apj, 768, 111

\bibitem[{{Shumko} {et~al.}(2014){Shumko}, {Gorceix}, {Choi}, {Kellerer},
  {Cao}, {Goode}, {Abramenko}, {Richards}, {Rimmele}, \&
  {Marino}}]{Shumko+etal2014SPIE.9148E..35S}
{Shumko}, S., {Gorceix}, N., {Choi}, S., {Kellerer}, A., {Cao}, W., {Goode},
  P.~R., {Abramenko}, V., {Richards}, K., {Rimmele}, T.~R., \& {Marino}, J.
  2014, in Society of Photo-Optical Instrumentation Engineers (SPIE) Conference
  Series, Vol. 9148, Adaptive Optics Systems IV, 914835

\bibitem[{{Skogsrud} {et~al.}(2016){Skogsrud}, {Rouppe van der Voort}, \& {De
  Pontieu}}]{Skogsrud+etal2016ApJ...817..124S}
{Skogsrud}, H., {Rouppe van der Voort}, L., \& {De Pontieu}, B. 2016, \apj,
  817, 124

\bibitem[{{Solanki}(2003)}]{Solanki2003A&ARv..11..153S}
{Solanki}, S.~K. 2003, \aapr, 11, 153

\bibitem[{{Solanki} \& {Montavon}(1993)}]{Solanki+Montavon1993A&A...275..283S}
{Solanki}, S.~K., \& {Montavon}, C.~A.~P. 1993, \aap, 275, 283

\bibitem[{{Stanchfield} {et~al.}(1997){Stanchfield}, {Thomas}, \&
  {Lites}}]{stanchfield+thomas+lites1997}
{Stanchfield}, D.~C.~H., {Thomas}, J.~H., \& {Lites}, B.~W. 1997, \apj, 477,
  485

\bibitem[{{Tian} {et~al.}(2014{\natexlab{a}}){Tian}, {DeLuca}, {Reeves},
  {McKillop}, {De Pontieu}, {Mart{\'{\i}}nez-Sykora}, {Carlsson}, {Hansteen},
  {Kleint}, {Cheung}, {Golub}, {Saar}, {Testa}, {Weber}, {Lemen}, {Title},
  {Boerner}, {Hurlburt}, {Tarbell}, {Wuelser}, {Kankelborg}, {Jaeggli}, \&
  {McIntosh}}]{Tian+etal.Osci.2014ApJ...786..137T}
{Tian}, H., {DeLuca}, E., {Reeves}, K.~K., {McKillop}, S., {De Pontieu}, B.,
  {Mart{\'{\i}}nez-Sykora}, J., {Carlsson}, M., {Hansteen}, V., {Kleint}, L.,
  {Cheung}, M., {Golub}, L., {Saar}, S., {Testa}, P., {Weber}, M., {Lemen}, J.,
  {Title}, A., {Boerner}, P., {Hurlburt}, N., {Tarbell}, T.~D., {Wuelser},
  J.~P., {Kankelborg}, C., {Jaeggli}, S., \& {McIntosh}, S.~W.
  2014{\natexlab{a}}, \apj, 786, 137

\bibitem[{{Tian} {et~al.}(2014{\natexlab{b}}){Tian}, {Kleint}, {Peter},
  {Weber}, {Testa}, {DeLuca}, {Golub}, \&
  {Schanche}}]{Tian+etal.BD.2014ApJ...790L..29T}
{Tian}, H., {Kleint}, L., {Peter}, H., {Weber}, M., {Testa}, P., {DeLuca}, E.,
  {Golub}, L., \& {Schanche}, N. 2014{\natexlab{b}}, \apjl, 790, L29

\bibitem[{{Tiwari} {et~al.}(2016){Tiwari}, {Moore}, {Winebarger}, \&
  {Alpert}}]{Tiwari+etal2016ApJ...816...92T}
{Tiwari}, S.~K., {Moore}, R.~L., {Winebarger}, A.~R., \& {Alpert}, S.~E. 2016,
  \apj, 816, 92

\bibitem[{{Tsuneta} {et~al.}(2008){Tsuneta}, {Ichimoto}, {Katsukawa},
  {et~al.}}]{Tsuneta+etal2008SoPh..249..167T}
{Tsuneta}, S., {Ichimoto}, K., {Katsukawa}, Y., {et~al.} 2008, \solphys, 249,
  167

\bibitem[{{Vissers} {et~al.}(2015){Vissers}, {Rouppe van der Voort}, \&
  {Carlsson}}]{Vissers+etal2015ApJ...811L..33V}
{Vissers}, G.~J.~M., {Rouppe van der Voort}, L.~H.~M., \& {Carlsson}, M. 2015,
  \apjl, 811, L33

\bibitem[{{Wang} {et~al.}(2012){Wang}, {Deng}, \&
  {Liu}}]{Wang+Deng+Liu2012ApJ...748...76W}
{Wang}, H., {Deng}, N., \& {Liu}, C. 2012, \apj, 748, 76

\bibitem[{{Wiehr} \&
  {Stellmacher}(1989)}]{Wiehr+Stellmacher1989A&A...225..528W}
{Wiehr}, E., \& {Stellmacher}, G. 1989, \aap, 225, 528

\bibitem[{{W{\"o}ger} \& {von der
  L{\"u}he}(2007)}]{Woger+Luhe2007ApOpt..46.8015W}
{W{\"o}ger}, F., \& {von der L{\"u}he}, O. 2007, \ao, 46, 8015

\bibitem[{{Yurchyshyn} {et~al.}(2015){Yurchyshyn}, {Abramenko}, \&
  {Kilcik}}]{Yurchyshyn+etal2015ApJ...798..136Y}
{Yurchyshyn}, V., {Abramenko}, V., \& {Kilcik}, A. 2015, \apj, 798, 136

\end{thebibliography}
\end{document}